\documentclass[english]{article}
\usepackage[T1]{fontenc}
\usepackage[latin9]{inputenc}
\usepackage[letterpaper]{geometry}
\usepackage{color}
\usepackage{graphicx}
\usepackage{amssymb}
\usepackage{babel}

\begin{document}

\title{Numerical deflation of beach balls with various Poisson's ratios:
from sphere to bowl's shape}

\author{Catherine Quilliet}

\maketitle

\begin{abstract}
We present a numerical study of the shape taken by a spherical elastic
surface when the volume it encloses is decreased. For the range of
2D parameters where such a surface may model a thin shell of an isotropic
elastic material, the mode of deformation that develops a single depression
is investigated in detail. It occurs via buckling from sphere toward
an axisymmetric dimple, followed by a second buckling where the depression
loses its axisymmetry through folding along portions of meridians.
For the thinnest shells, a direct transition from the spherical conformation
to the folded one can be observed. We could exhibit unifying master
curves for the relative volume variation at which first and second
buckling occur, and clarify the role of the Poisson's ratio. In the
folded conformation, the number of folds and inner pressure are investigated,
allowing us to infer shell features from mere observation and/or knowledge
of external constraints.
\end{abstract}

\section{Introduction\label{sec:Introduction}}

Let's consider a thin shell of an elastic isotropic material, such
as a beach ball, and deflate it. What would be its shape ?

This question is not restricted to garrulous familial shores : fundamental
and applied physics nowadays presents legions of easily deformable
soft objects, and knowing what governs their shapes gives the powerful
possibility of inferring mechanical properties from simple observations,
without contact. Among these deformable objects, an increasing number
derives from spherical symmetry, that is omnipresent at scales where
surface effects overcome volume forces such as gravity. The numerical
study presented in this paper discusses the shapes taken by spherical
thin shells of isotropic materials when their inner volume is decreased
by a significant amount. Such a systematic and quantitative study
will help deciphering conformations observed in \emph{e.g.} Soft Matter
(lock-and-key colloids\cite{Sacanna2011}, multiwall capsules\cite{SunNonCollapsed11},
particles design through evaporation\cite{Tsapis2005,Bahadur2012}),
galenics (encapsulation \cite{Delcea2011}), microfluidics (microtanks
\cite{Kiryukhin2011}) or medicine (ultrasound contrast agents\cite{JASA2011}),
under the action of an external pressure or other possibly isotropic
fields such as concentration in evaporation/dissolution phenomena,
which also shapes objects in Nature \cite{Katifori2010}.

When an elastic spherical shell has its inner volume lowered, it first
deforms through in-plane compression that respects the spherical symmetry.
Then it undergoes a symmetry breaking in order to relax a high stretch
energy into much lower bending energy, by reversion of a spherical
cap (creation of an axisymmetric depression, or {}``dimple''). The
onset of this sudden transition, or buckling, under external pressure
was studied long ago by Pogorelov and Landau\cite{Pogorelov,Landau}.
They showed that the dimple should nucleate over a critical outside/inside
pressure difference scaling like $Y_{3D}\left(\frac{d}{R}\right)^{2}$
(in what follows, we will refer to this latter quantity by $\Delta P_{Landau}$),
where $d$ is the shell thickness, $R$ its radius and $Y_{3D}$ the
Young modulus of the material that makes it up. Its edge (or {}``rim'')
has a transversal extension $\sqrt{dR}$. One of the key of their
calculation being the assumption that buckling occurs for dimples
such that maximum deflection is of order $d$, spherical geometry
imposes then that $\sqrt{dR}$ is also the radius of the dimples that
forms. Besides, classical buckling analysis provided dependence of
the buckling pressure with the Poisson's ratio\cite{Hutchinson67,Pogorelov,Landau}.

Results focusing on deformations through further deflation are, mainly,
more recent. A few months ago, stability analysis allowed a detailed
study of buckling toward axisymmetric conformations\cite{Knoche2011}.
Experimental \cite{Carlson67,Tsapis2005,Zoldesi2005,Datta2010,Sen2010,Sacanna2011,Quemeneur11}
and numerical \cite{Tsapis2005,Vliegen2011,Quemeneur11} deflation
studies showed shapes holding several dimples, also called {}``multiple
indentation''. These conformations compete with experimental observations
of shapes holding a single depression\cite{Zoldesi2005,Sen2010},
possibly losing axisymmetry\cite{Okubo2001} or exhibiting folding
perpendicularly to the rim \cite{Quilliet 2008,Datta2010}. Similar
shapes are observed in shells under a point load \cite{Pauchard98,Vaziri2009}
or pressed against a wall \cite{Komura 2005,Vaziri2009}. Secondary
buckling by folding of the single depression, also called {}``polygonal
indentation'', under isotropic constraint was numerically retrieved
with surface models\cite{Quilliet 2008,Vliegen2011}. Thin shells
with a single depression, either axisymmetric or polygonal, appear
to present a conformation of lower energy than the metastable multiple
indentation\cite{Quilliet 2008,Vliegen2011,Quemeneur11}. In the case
of an axisymmetric dimple, this can be easily understood since elastic
energy mainly concentrates in dimple edges as bending energy, with
an energy per edge length that weakly varies with dimple size. Hence
dimples coalescence lowers the total elastic energy \cite{NoteSurCQ2006}.
Nevertheless, more than one dimple may nucleate i\textcolor{black}{f
the deflation is rapid enough, leading to metastable\cite{Tsapis2005,Vliegen2011,Quemeneur11}
multi-indented shapes. The term {}``rapid'' is to be taken on a
wide acception here. Experimentally, it may correspond to situations
where dissipation (due to material viscosity or to fluid flows accompanying
the deformation) prevents dimple growth, which favors secondary nucleation
once $\Delta P_{Landau}$ is reached, and where subsequent kinetics
prevents thermally activated coalescence between adjacent dimples.
Numerically, minimization may reproduce such metastable situations
\cite{Tsapis2005,Vliegen2011,Quemeneur11}, since (i) large volume
increments favors the creation of extra dimples, by making difficult
to find the cooperative displacement of vertices that corresponds
to rim rolling in dimple growth }(ii) depending on the way curvatures
are calculated, energy barriers that prevents from dimple coalescence
may be overcome or not.\textcolor{black}{{} For {}``slow'' deflations,
a single dimple can appear and grow, or freshly nucleated dimples
may coalesce into a single one. Such {}``slow'' deflation provides
an axisymmetric bowl-like shape, that may undergo under further deflation
a transition toward a non-axisymmetric depression, }\textcolor{black}{\emph{i.e.}}\textcolor{black}{{}
polygonal indentation\cite{Quilliet 2008,Vliegen2011}.}

\textcolor{black}{We present here a systematic numerical study of
such {}``slow'' def}lations leading to shapes with a single depression.
In this purpose, we used a surface model taking into account recent
developments, presented in section \ref{sec:Surface-model}. We clearly
expose the correspondence between 2D parameters of the model surface,
and 3D properties of the real object of nonzero thickness, expliciting
the role of the different significative parameters. Particular emphasis
is put on a parameter often underconsidered: the Poisson's ratio.

The whole study allows to determine parameters of importance for the
transitions sphere $\rightarrow$ axisymetric bowl (section \ref{sec:First-order-transition})
and axisymmetric bowl $\rightarrow$ polygonal indentation (section
\ref{sec:Second-order-transition}), both for the detailed shape in
polygonal indentation, and for inner pressure. Furthermore, we took
particular care to provide empirical dependence laws for practical
use.

\section{Surface model\label{sec:Surface-model}}

\textcolor{black}{Surface model, where out-of-plane and in-plane deformations
are formally uncoupled, is for long considered as valid to describe
the deformation of thin sheets (plates or shells) \cite{Landau,BookAudoly}.
For thin sheets without spontaneous curvature (}\textcolor{black}{\emph{i.e}}\textcolor{black}{.
an elementary surface portion of the sheet, freed from constraints
exerted by surrouding material, remains flat at equilibrium), the
energies per surface unit that are to be considered in this surface
model are of two kinds: firstly, a curvature term that can express
$\frac{1}{2}\kappa c^{2}+\overline{\kappa}g$ \cite{Helfrich 1973,Landau},
where $c=\frac{1}{R_{1}}+\frac{1}{R_{2}}$ and $g=\frac{1}{R_{1}}\times\frac{1}{R_{2}}$
are respectively the mean and Gaussian curvatures ($R_{1}$ and $R_{2}$
being the local principal curvature radii), and $\kappa$ and $\overline{\kappa}$
are respectively the mean and Gaussian curvature constants\cite{NoteCourbure}.
The other term may be written, in a Hookean linear model: $\frac{1}{2}\epsilon_{ij}K_{ijkl}\epsilon_{kl}$,
where $\epsilon_{ij}$ and $K_{ijkl}$ respectively represent the
two-dimensional strain and elasticity tensors for in-plane deformations.
For an homogeneous and isotropic surface, the nonzero terms of the
two-dimensional elasticity tensor are $K_{xxxx}=K_{yyyy}=\frac{Y_{2D}}{1-\nu_{2D}^{2}}$,
$K_{xxyy}=K_{yyxx}=\frac{\nu_{2D}\, Y_{2D}}{1-\nu_{2D}^{2}}$ and
$K_{xyxy}=K_{yxyx}=\frac{Y_{2D}}{1+\nu_{2D}}$, with $Y_{2D}$ the
two-dimensional Young modulus and $\nu_{2D}$ the two-dimensional
Poisson ratio, which is comprized between -1 and 1 \cite{Landau}.
This in-plane elasticity term can be rewritten as $\frac{Y_{2D}}{2(1+\nu_{2D})}\left[\mbox{Tr}(\epsilon^{2})+\frac{\nu_{2D}(\mbox{Tr}\epsilon)^{2}}{1-\nu_{2D}}\right]$
for the sake of concision.}

In a linear approximation, the relation between the 2D parameters
and the 3D features of the plate (Young modulus $Y_{3D}$, Poisson's
ratio $\nu_{3D}$, thickness $d$) with zero boundary tangential constraints
\cite{Landau} is expressed as (detailed \emph{e.g.} in \cite{Komura 2005}
or \cite{JASA2011}):\begin{equation}
\nu_{2D}=\nu_{3D}=\nu\label{eq:nu}\end{equation}
\begin{equation}
Y_{2D}=Y_{3D}d\label{eq:Y2D}\end{equation}
\begin{equation}
\kappa=\frac{Y_{3D}}{12\left(1-\nu^{2}\right)}\, d^{3}\label{eq:kappa}\end{equation}
\begin{equation}
\overline{\kappa}=\left(\nu-1\right)\kappa=-\frac{Y_{3D}}{12\left(1+\nu\right)}\, d^{3}\label{eq:kappabarre}\end{equation}

Since for bulk materials the maximum value of $\nu_{3D}$ is $\frac{1}{2}$
for thermodynamic reasons \cite{Landau}, one can notice that the
range of 2D Poisson's ratio that effectively describes a thin plate
of an isotropic material is limited to a maximum value of $\frac{1}{2}$.
In other terms, even a thin plate of an incompressible isotropic material
cannot behave as an incompressible surface (where\emph{ }$\nu_{2D}=1$),
thanks to the possibility of having its thickness varied. On the other
limit, Poisson's ratio can reach -1 as a lower value, but negative
values correspond to less common {}``auxetic'' materials.

Conversely the thickness of the plate, as a function of 2D parameters,
writes:\begin{equation}
d=\sqrt{12\left(1-\nu^{2}\right)\frac{\kappa}{Y_{2D}}}\label{eq:dsurR}\end{equation}

\textcolor{black}{For describing surfaces with asymmetric properties,
the notion of {}``spontaneous curvature'' was introduced by W. Helfrich
\cite{Helfrich 1973}. It was recently shown that to describe the
deformations of an initially stress-free thin shell of radius $R$,
the three contributions (in-plane, mean curvature and gaussian energy,
that may have a non-vanishing part even for closed surfaces, depending
on the definition of a nonzero spontaneous curvature) can be rewritten
in an easily computable way as \cite{MarmottantSM2011}: \begin{equation}
E_{elastic}=cst\,+\,\int_{shell\ surface}\left[\frac{1}{2}\kappa\left(c-c_{0}^{*}\right)^{2}+\frac{1}{2}\epsilon_{ij}K_{ijkl}\epsilon_{kl}+\gamma_{eff}\right]\, dS\label{eq:Elastic energy}\end{equation}
with $c_{0}^{*}=\frac{1+\nu}{R}$ being the effective spontaneous
curvature, and $\gamma_{eff}=\frac{\left(1-\nu^{2}\right)\kappa}{2R^{2}}$
an effective surface tension.}

\textcolor{black}{This expression is slightly different from the one
used}\textcolor{black}{\emph{ }}\textcolor{black}{in \cite{Quilliet 2008},
hence we will quantitatively discuss, in the results, modifications
induced by the use of this more complete expression.}

More generally, we will consider the influence of sphere size through
the use of the adimensionalized Föppl-von K\'arm\'an number\cite{Lidmar2003}
$\gamma=\frac{Y_{2D}R^{2}}{\kappa}$, that gives the order of magnitude
of the ratio between in-plane and out-of plane deformation energies\cite{Komura 2005}.
An elastic surface with the energy given in equation (\ref{eq:Elastic energy})
can effectively describe a thin shell of an isotropic material if
$12\left(1-\nu^{2}\right)/\gamma\ll1$, in addition to the condition
$\nu_{2D}\leq\frac{1}{2}$. In this range, $\gamma$ roughly scales
like $\left(\frac{R}{d}\right)^{2}$. Out of this range, such a surface
model does not correspond to any thin shell of an isotropic material
; it can nevertheless describe different types of objects, \emph{e.g.}
gel-phase vesicles \cite{Quemeneur11} that can hence be considered
as thin shells of non isotropic materials.

Numerical experiments are performed by minimizing the elastic energy
as expressed in equation (\ref{eq:Elastic energy}) for different
inner volumes, with the free software Surface Evolver\cite{Brakke}.
A whole in-silico deflation experiment (from $V_{init}=\frac{4}{3}\pi R^{3}$
to $\approx0.2\times V_{init}$) is realized through a succession
of different equilibrium states, these latter found according the
process described in \cite{Quilliet 2008}, and calculated successively
for inner volumes decreased by steps of at maximum 2\% of the initial
volume (steps amplitude is reduced in some situations in order to
avoid the nucleation of secondary dimples).

\section{First-order transition toward axisymmetric depression\label{sec:First-order-transition}}

Deflation of a spherical elastic surface, at imposed either volume
or external pressure, causes an abrupt buckling from the spherical
conformation in order to release in-plane compressive stress. The
purpose here is to compare the numerical approach described in the
previous paragraph to known features of this buckling, for the range
of parameters that scans the generality of thin shells of isotropic
material.

\subsection{Buckling pressure}

\begin{figure}
\begin{center}\includegraphics{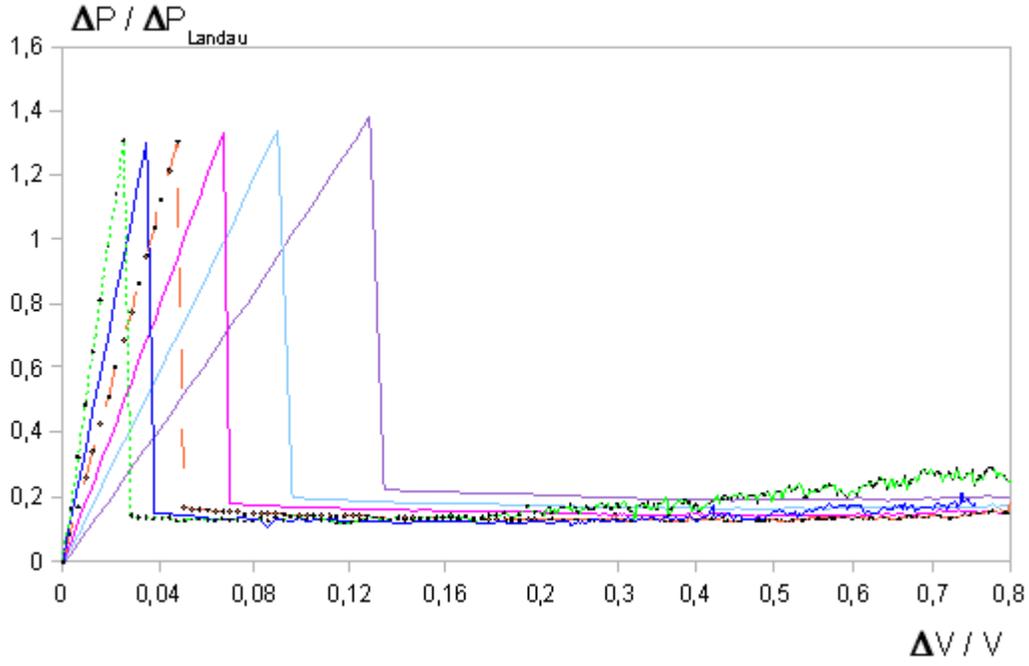}\end{center}

\caption{\label{fig:Pressure-difference-}Pressure difference $\Delta P=P_{ext}-P_{int}$
adimensionalized by $\Delta P_{Landau}=Y_{3D}\left(\frac{d}{R}\right)^{2}$
(or $\left[12\, Y_{2D}\kappa\left(1-\nu^{2}\right)\right]^{1/2}/R^{2}$
in 2D parameters), as a function of the relative volume variation,
for surfaces of similar Poisson's ratio $\nu=-0.5$. Dotted green:
$\gamma=1.17\times10^{5}$ ($\frac{d}{R}=8.8\times10^{-3}$); blue:
$\gamma=6.07\times10^{4}$ ($\frac{d}{R}=1.22\times10^{-2}$), interrupted
orange: $\gamma=3.22\times10^{4}$ ($\frac{d}{R}=1.67\times10^{-2}$),
magenta: $\gamma=1.17\times10^{4}$ ($\frac{d}{R}=2.31\times10^{-2}$),
light blue: $\gamma=9.33\times10^{3}$ ($\frac{d}{R}=3.11\times10^{-2}$),
parma: $\gamma=4.67\times10^{3}$ ($\frac{d}{R}=4.39\times10^{-2}$).
Notice scale switch at $\frac{\Delta V}{V}=0.2$. Points (corresponding
each to a minimization) emphasizes two curves with typical behaviour
after buckling: increasing (green curve), and plateauing (orange)
.}

\end{figure}

First buckling relaxes in-plane constraints, and causes drastic drop
of the inside/outside pressure difference $\Delta P=P_{ext}-P_{int}$.
Figure \ref{fig:Pressure-difference-} displays typical evolutions
for the pressure difference: first a linear increase followed by a
drop at first buckling, after which pressure difference varies in
a much lesser extent. Linear behaviour is expected before the first
buckling due to the relation between pressure and elastic energy $P_{ext}-P_{int}=\partial E_{elastic}/\partial\left(\Delta V\right)$
(detailed in \cite{JASA2011}), and quadratic dependence of $E_{elastic}$
with $\Delta V=V_{init}-V$ (see \emph{e.g.} \cite{Quilliet 2008}).

\begin{figure}
\begin{center}\includegraphics[width=10cm]{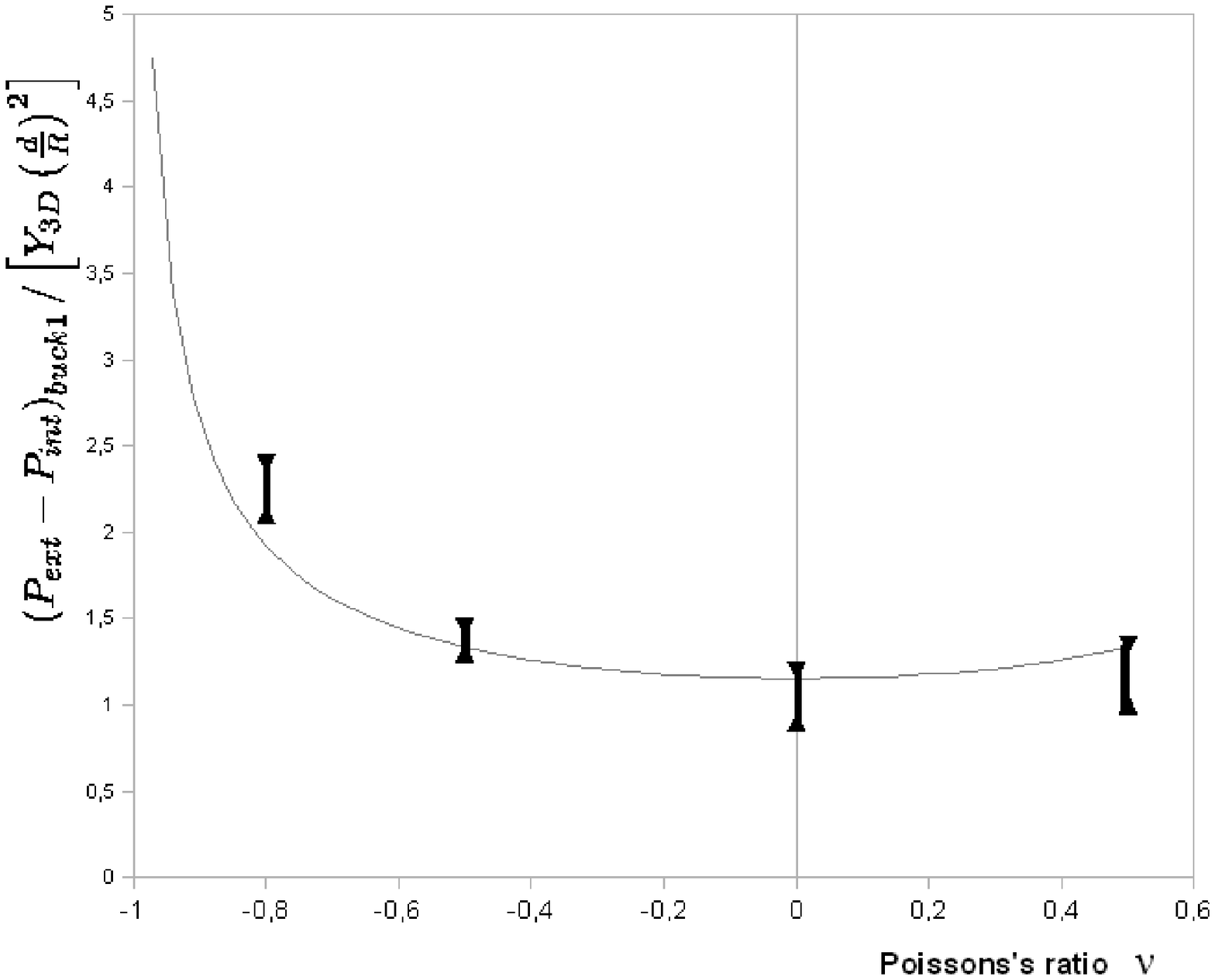}\end{center}\caption{\label{fig:Hutch}Pressure $\left(\Delta P\right)_{buck\,1}=\left(P_{ext}-P_{int}\right)_{buck\,1}$
at first sphere$\rightarrow$bowl buckling, adimensionized by $\Delta P_{Landau}$.
Error bars represent the range of this quantity for $\gamma$ between
$4.67\times10^{3}$ and $2.33\times10^{5}$. Line: theoretical value
$2\left[3\left(1-\nu^{2}\right)\right]^{-1/2}$, from equation \ref{eq:PressionBuckHutchinson}.}

\end{figure}

In the simulations presented here, designed not to be stuck in multi-indentation
conformations of higher energy, or other less stable, the first buckling
leads to a single axisymmetric dimple. It is expected to happen when
the external overpressure $\Delta P=P_{ext}-P_{int}$ reaches the
critical value \cite{Hutchinson67,Knoche2011}:\begin{equation}
\Delta P_{c}=2\left[3\left(1-\nu^{2}\right)\right]^{-1/2}\times Y_{3D}\left(\frac{d}{R}\right)^{2}\label{eq:PressionBuckHutchinson}\end{equation}
As expected, first buckling in our simulations effectively occurs
at a pressure difference of order $\Delta P_{Landau}=Y_{3D}\left(\frac{d}{R}\right)^{2}$
(section \ref{sec:Introduction}). For a given Poisson's ratio, the
incertitude due to discrete volume increments does not allow to conclude
that $\Delta P_{buckling1}/\Delta P_{Landau}$ is affected by $\gamma$
(see \emph{e.g.} figure \ref{fig:Pressure-difference-} displaying
several cases at $\nu=-0.5$ ). The effect of the Poisson's ratio
is displayed on figure \ref{fig:Hutch}: it shows that the pressure
which induces buckling in our simulations quantitatively follows the
theoretical equation \ref{eq:PressionBuckHutchinson}, which reinforces
the validity of our approach.

\subsection{Buckling volume}

The mechanism of buckling in axisymmetric conformations was quite
recently investigated in detail by Knoche \emph{et al}\cite{Knoche2011},
with the study of various metastability branches. At first significative
order, their calculations show that for the trivial isotropic ({}``spherical'')
deformation, the relative volume variation due to an external overpressure
$\Delta P=P_{ext}-P_{int}$ expresses:\[
\frac{\Delta V}{V}=\frac{3\left(1-\nu\right)}{2}\times\frac{R}{d}\times\frac{\Delta P}{Y_{3D}}\]

Hence the deflation at buckling pressure writes\cite{Hutchinson67,Knoche2011}:

\begin{equation}
\left(\frac{\Delta V}{V}\right)_{buck\,1}=\sqrt{3\left(\frac{1-\nu}{1+\nu}\right)}\times\frac{d}{R}\label{eq:Buck1 tridi}\end{equation}

\begin{figure}
\begin{center}\includegraphics[scale=0.5]{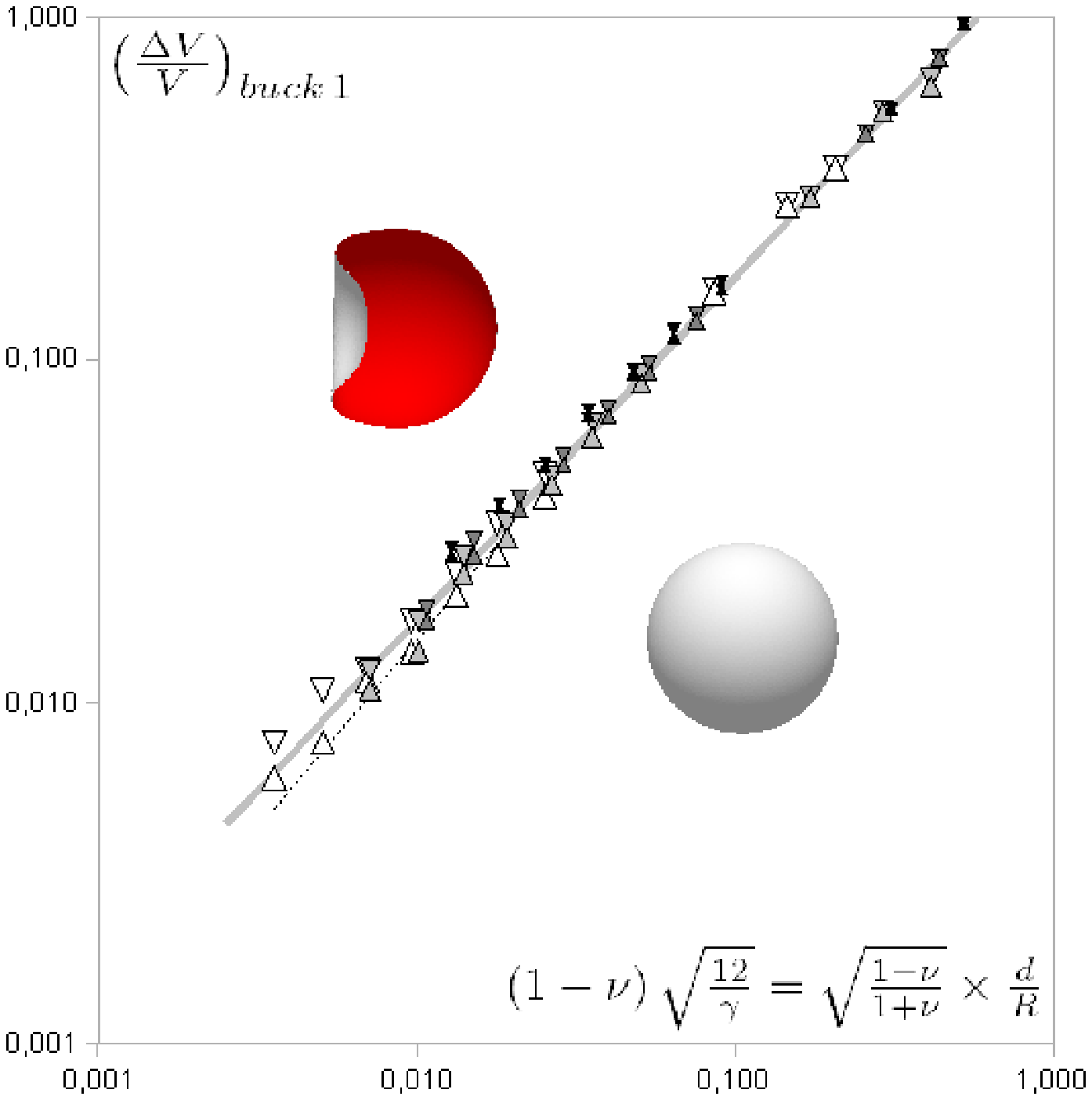}\end{center}

\caption{\label{fig:SeuilsB1}Relative volume variation $\left(\frac{\Delta V}{V}\right)_{buck\,1}$
at the sphere$\rightarrow$bowl buckling (lower and upper limit of
the error bars are indicated respectively by upward and downmard triangles).
White: $\nu=0.5$ ; light gray: $\nu=0$ ; dark grey: $\nu=-0.5$
; black: $\nu=-0.8$. Grey line: $\left(\frac{\Delta V}{V}\right)_{buck\,1}=6\left(1-\nu\right)\gamma^{-\frac{1}{2}}$
(this is equation \ref{eq:Buck1 tridi} expressed in 2D parameters).
Dotted line: relation established in \cite{Quilliet 2008} for $\nu=1/3$:
taking into account the gaussian curvature elastic energy increases
volume threshold by at maximum $20\%$ for the thinnest shells of
the shells previously studied. Illustrative inserts display deflated
spherical surfaces at $\gamma=4666$ and $\nu=-0.8$ (hence $\frac{d}{R}=0.0304$),
with relative volume variations respectively $\frac{\Delta V}{V}=0.161$
(spherical) and 0.167 (buckled, section view, same scale).}

\end{figure}

Figure \ref{fig:SeuilsB1} presents values of $\left(\frac{\Delta V}{V}\right)_{buck\,1}$
from numerical simulations, as a function of a combination of $\gamma$
and $\nu$ translated in 3D parameters: here also the theoretical
equation \ref{eq:Buck1 tridi} is quantitatively retrieved. One may
notice (since by essence $\left(\frac{\Delta V}{V}\right)_{buck\,1}\leq1$)
that the sphere$\rightarrow$bowl transition vanishes for $\frac{d}{R}\geq\sqrt{\left(1+\nu\right)/3\left(1-\nu\right)}$,
which indicates a destabilization of the axisymmetric bowl for the
most auxetic materials. This was qualitatively expected from the $\kappa$
divergence in this limit, which makes curvature deformations prohibitive
compared to in-plane compressions\cite{NoteKhi}.

Using the 2D parameters of the surface model, relation \ref{eq:Buck1 tridi}
also writes:\begin{equation}
\left(\frac{\Delta V}{V}\right)_{buck\,1}=6\left(1-\nu\right)\gamma^{-\frac{1}{2}}\label{eq:Buck1 bidi}\end{equation}

Extending our purpose out of the range of 2D parameters that effectively
can describe a thin shell of isotropic material, we may remark that
extrapolation to $\nu=1$ induces vanishing of $\left(\frac{\Delta V}{V}\right)_{buck\,1}$,
and hence destabilization of the spherical conformation : this limit
corresponds to incompressible surfaces, that will necessarily undergo
a deformation implying curvature even for the smallest volume decreases,
since area variation is prohibited. For the thickest shells, \emph{i.e.}
$\gamma\lesssim80$, shape transition occurs not any more through
sudden inversion of a spherical cap, but by slowly deforming into
an ovoid, that flattens at the location of future depression under
further deflation. We did not, here, specifically study this extreme
behaviour.

For practical purposes, we may notice that the large range of parameters
explored shows that in terms of volume, the onset of buckling mainly
depends on the relative thickness $\frac{d}{R}$, with only a weak
influence of the Poisson's ratio. Since this latter ranges between
0 and $\frac{1}{2}$ for common materials (\emph{i.e.} non auxetic,
with $\nu\geq0$), the prefactor of $\frac{d}{R}$ varies between
$\sqrt{3}$ and 1, which is much narrower than the range in $\frac{d}{R}$
that can be explored.

\section{Second-order transition toward polygonal depression\label{sec:Second-order-transition}}

\subsection{Location of the transition}

\begin{figure}
\begin{center}\includegraphics[width=3.9cm]{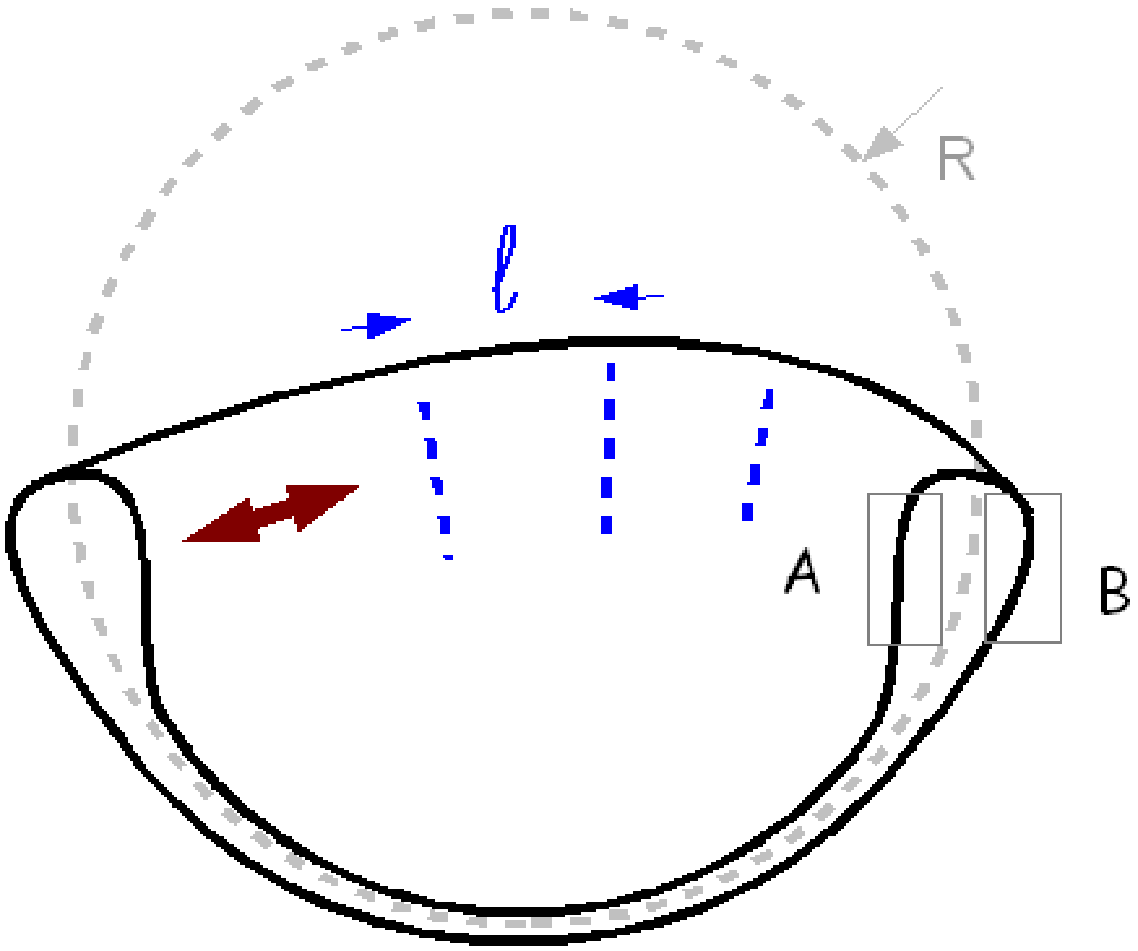}\includegraphics[width=0.2cm]{PetitBlanc}\includegraphics[width=3.9cm]{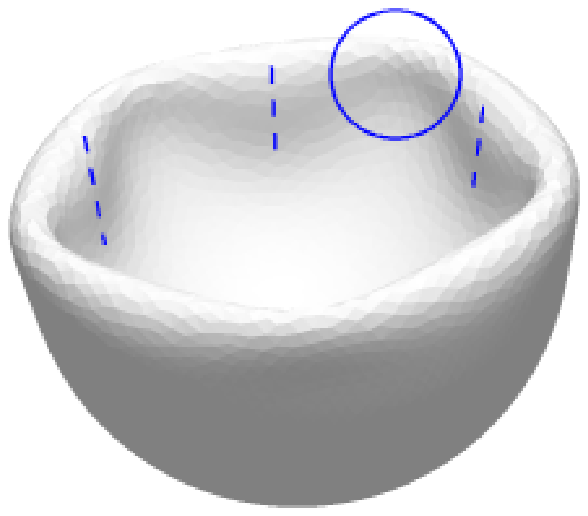}\end{center}

\caption{\label{fig:SchemCaps}Left: schematisation of a spherical surface
near complete deflation. Inner part of the rim endures a compressive
stress in the direction indicated by the red double arrow. Relaxation
of in-plane deformation occurs, for the thinnest shells, via undulation
deformation, generating wrinkles (folds) all along the rim (examples
schematized with blue interrupted lines). Length $\ell$ stands for
lateral extension of the wrinkles. Right: simulation with $\gamma=9.33\times10^{3}$,
$\nu=0.5$ ($\frac{d}{R}=0.031$) and $\frac{\Delta V}{V}=0.562$
). Maxima of undulation are stressed with blue interrupted lines ;
circle locates the zone of high curvature that forms the apex of the
s-cone.}

\end{figure}

\begin{figure}
\begin{center}\includegraphics[width=10cm]{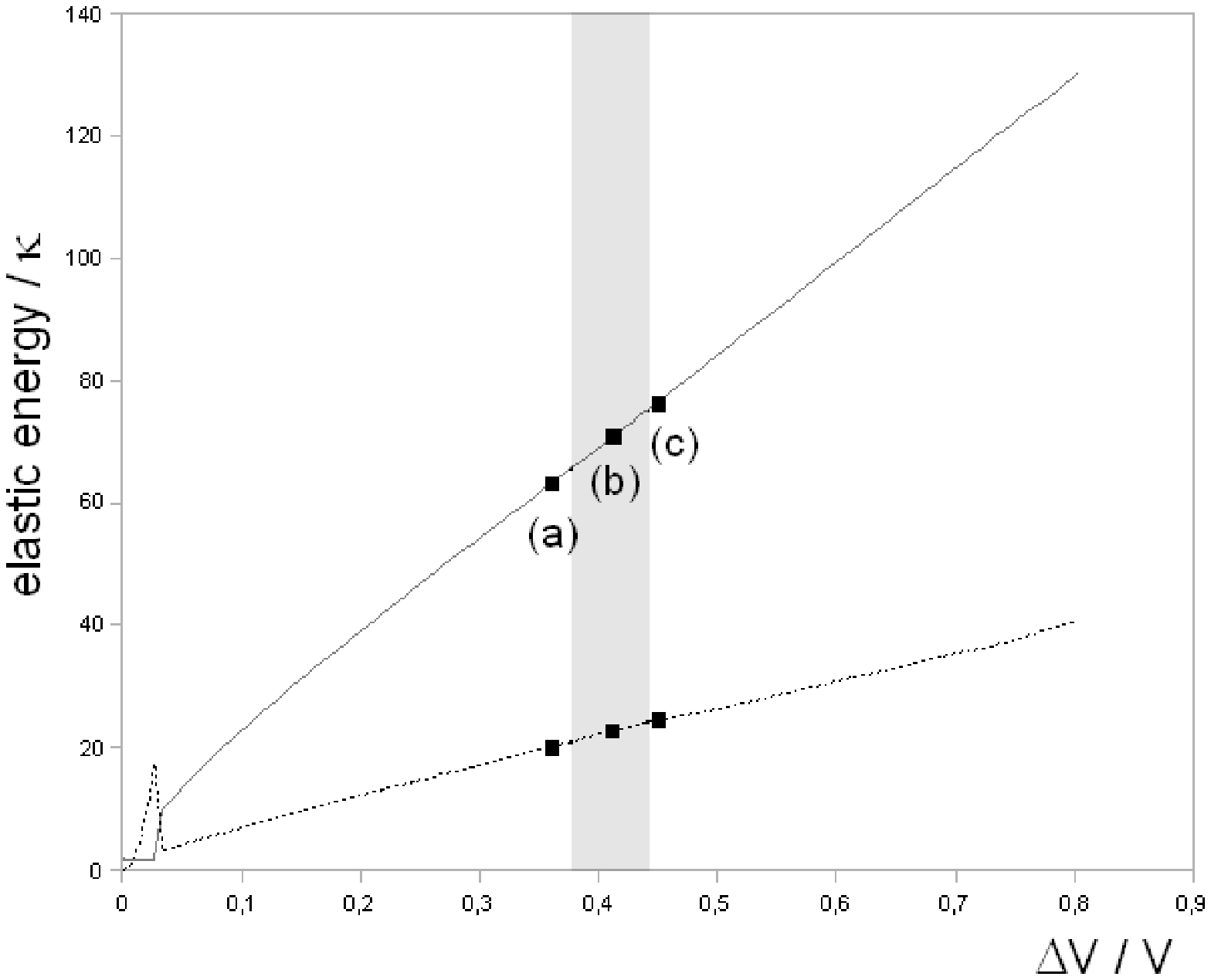}

\includegraphics[width=2.5cm]{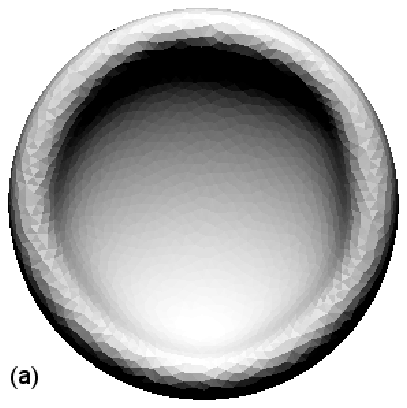}\includegraphics[width=2.5cm]{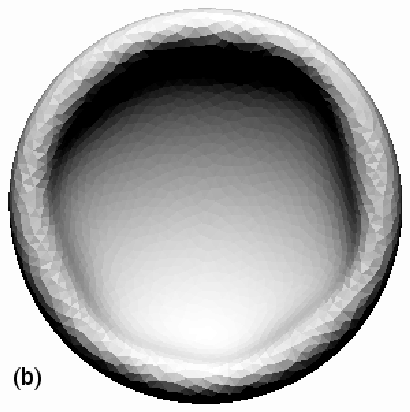}\includegraphics[width=2.5cm]{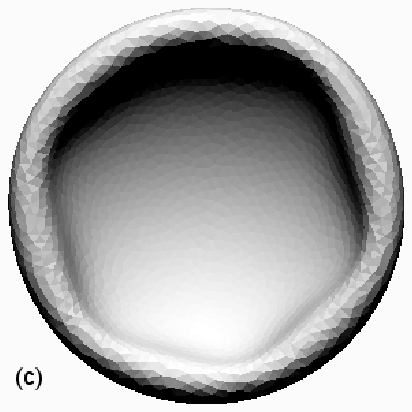}\end{center}

\caption{\label{fig:DefEnergies}Deformation energies at different relative
volume variations (volume step $\approx0.6\%$ of the initial volume),
adimensionalized by the curvature constant $\kappa$, for $\gamma=9.33\times10^{3}$
and $\nu=0.5$ ($\frac{d}{R}=0.031$). Interrupted line: Hookean in-plane
deformation energy, linked to the second significative term of eq.
(\ref{eq:Elastic energy}), \emph{i.e.} $\frac{1}{2}\epsilon_{ij}K_{ijkl}\epsilon_{kl}$.
Grey continuous line: elastic energy linked to the first significative
term :\emph{ }$\frac{1}{2}\kappa\left(c-c_{0}^{*}\right)^{2}$. This
term does not vanish at $\frac{\Delta V}{V}=0$ because $c_{0}^{*}\neq c_{0}$
; in the total elastic energy, it is counterbalanced by the $cst$
and the effective surface term expressed by $\gamma_{eff}$ in equation
(\ref{eq:Elastic energy}). Effective surface energy (not represented
here) varies at maximum by $-0.33\,\kappa$ , and on an amplitude
$0.03\,\kappa$ in the nonspherical conformations. Grey zone indicates
the second transition from axisymmetric bowl shape to depression with
inner wrinkles ({}``polygonal indentation''\cite{Vliegen2011}),
determined as explained in the text. First transition from sphere
to axisymmetric bowl occurs through abrupt decrease of the in-plane
deformation energy, at $\frac{\Delta V}{V}=0.03$.}

\end{figure}

In the axisymmetric bowl shape, global bending of the rim on the equator
costs in-plane deformation: extension on the outer side of the rim,
and compression on the inner one. For the thinnest shells, compressive
stress parallel to the equator leads to a secondary buckling, where
the inner side of the rim undulates to adapt to axial compression
(fig. \ref{fig:SchemCaps}, left), forming folds, or {}``wrinkles'',
that deform the axisymmetric depression into a roughly polygonal shape
(fig. \ref{fig:SchemCaps}, right). \textcolor{black}{Such a conformation
mainly involves curvature deformations, much less energetic \cite{Komura 2005}
than compression energy that quadratically increases with $\frac{\Delta V}{V}$.}
Fig. \ref{fig:DefEnergies} shows how elastic energy dispatches between
in-plane and out-of-plane deformation energies in a typical numerical
deflation. Wrinkles match with the rim through a zone of high curvature
that has a folding role similar to what realizes the apex of d-cones
\cite{Pauchard97,Chaïeb98,Cerda2005}, except that the surface is
not developable but is spherical, hence the concept of {}``\emph{s}-cones''
proposed by Reis and Lazarus \cite{S-cones}.

\begin{figure}
\begin{center}\includegraphics{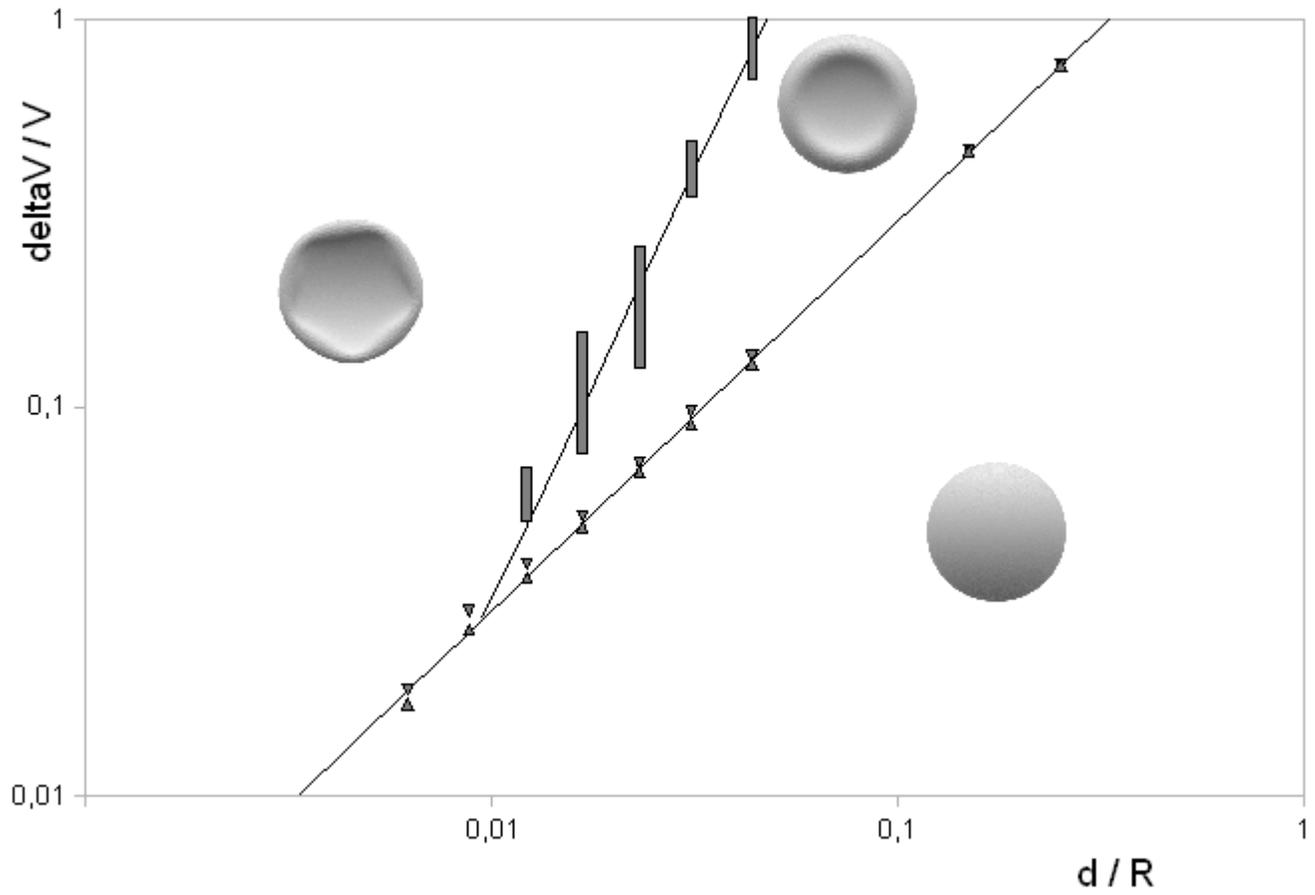}\end{center}

\caption{\label{fig:DgPhNuMsZero5}Shapes in the relative thickness / relative
volume variation phase diagra\textcolor{black}{m, for: $\nu=-0.5$.
Illustrations: shells with $d/R=0.031$ and $\nu=-0.5$ ($\frac{\Delta V}{V}$=0,
0.20 and 0.55).}}

\end{figure}

\begin{figure}
\begin{center}\includegraphics[scale=0.5]{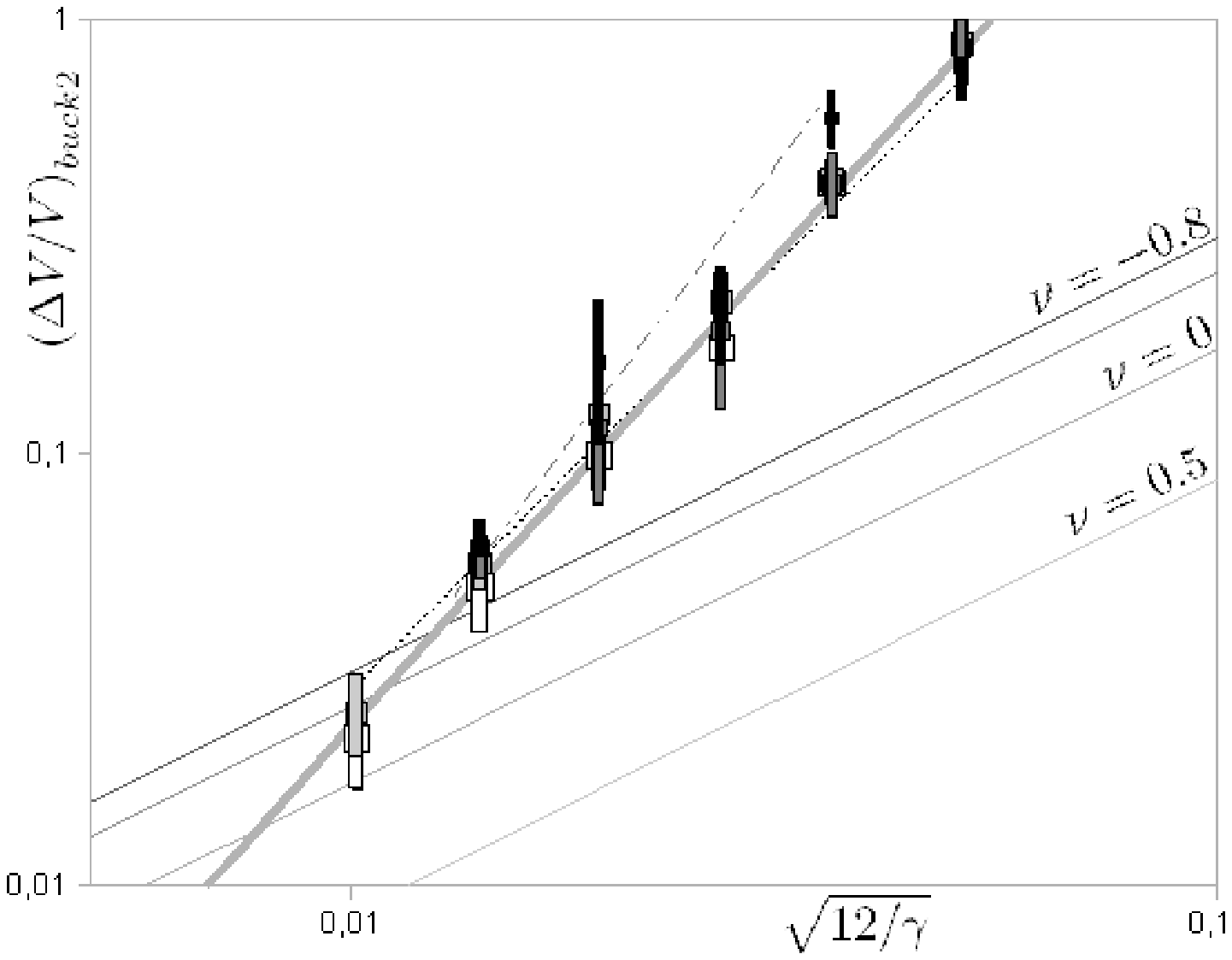}\end{center}

\caption{\label{fig:buck2}Points: relative volume variation at which the axisymmetric
depression becomes {}``polygonal'', versus $\sqrt{12/\gamma}$ (which
is to be identified with $d/R$ when $\nu=0$). White: $\nu=0.5$
; light grey: $\nu=0$ ; dark grey: $\nu=-0.5$ ; black: $\nu=-0.8$.
Wide grey line: equation \ref{eq:Buck2 tridi}. Continuous lines indicate
the location of the first-order buckling for different Poisson's ratio
(from equation \ref{eq:Buck1 tridi}), under which only the spherical
conformation is to be found. Non-continuous lines: in their domain
of validity, equations describing the axisymmetric/wrinkled bowl transition
published in previous works for $\nu=1/3$ (see text): dotted black
refers to \cite{Quilliet 2008}, interrupted grey to \cite{Vliegen2011}.}

\end{figure}

Secondary buckling from axisymmetric bowl shape to polygonal indentation,
quite smooth, is harder to detect than the first one (fig. \ref{fig:DefEnergies}).
The corresponding relative volume variation $\left(\frac{\Delta V}{V}\right)_{buck\,2}$
is determined on one hand by the maximum deflation before loss of
axisymmetry, and on the other hand by the $\frac{\Delta V}{V}$ at
which the rim presents convex zones under axial observation (as shown
on lower part of Fig. \ref{fig:DefEnergies}, subfigure c). Figure
\ref{fig:DgPhNuMsZero5} presents a typical shape phase diagram for
$\nu=-0.5$, with three distinct zones\textcolor{black}{: spherical
coformation, axisymmetric bowl and wrinkled depression. One may notice
that for the thinnest shells, the incremented deflation we numerically
performed shows a direct transition from the sphere to the wrinkled
bowl. Relative volume variation at first buckling nevertheless obeys
a single power-law on the whole range of relative thicknesses. We
focus in this section on the second buckling, hence considering only
deflations} where wrinkles appear on an already existing axisymmetric
conformation. Figure \ref{fig:buck2} displays how $\left(\frac{\Delta V}{V}\right)_{buck\,2}$
varies with the Föppl-von Karm\'an number $\gamma$ for different
Poisson's ratios. Data indicate a dependence in $\gamma$, but scattering
prevents from concluding on an influence by $\nu$ ; linear regression
on logarithms\cite{NoteChoixParamB2} provides with a correlation
coefficient of -0.99:\begin{equation}
\left(\frac{\Delta V}{V}\right)_{buck\,2}=8470\times\gamma^{-1.085}\label{eq:CourbMaitresseBuck2}\end{equation}

This expression is compatible with previous results obtained for $\nu=1/3$
with slightly different numerical models: the shell without gaussian
curvature evoked in section \ref{sec:Surface-model}\cite{Quilliet 2008},
and the spring model without spontaneous mean curvature of reference
\cite{Vliegen2011} (both are represented on figure \ref{fig:buck2}
for their range of validity).

Extrapolating equation \ref{eq:CourbMaitresseBuck2} up to $\frac{\Delta V}{V}=1$
suggests that this secondary buckling does not happen, \emph{i.e.}
single indentation keeps its axisymmetry, below a threshold value
$\gamma_{c,buck2}=4170$. In tridimensional parameters, relation (\ref{eq:CourbMaitresseBuck2})
expresses:\begin{equation}
\left(\frac{\Delta V}{V}\right)_{buck\,2}=571\times\left(\frac{d/R}{\sqrt{1-\nu^{2}}}\right)^{2.17}\label{eq:Buck2 tridi}\end{equation}

The dependence in $\nu$ for non-auxetic materials is even weaker
than for the axisymmetric buckling since it plays at maximum by a
factor 4/3. Besides, for the range of Poisson's ratio studied here,
there is no axisymmetric conformation to be expected for $d/R<0.003$.
Relation (\ref{eq:Buck2 tridi}) also implies that wrinkles are not
expected when $d/R\gtrsim0.054\times\sqrt{1-\nu^{2}}$ : a particular
consequence is that in wrinkles prevention, a very auxetic material
(with $\nu\rightarrow-1$) may help.

\subsection{Characterization of buckled shapes with the number of wrinkles}

\begin{figure}
\begin{center}\includegraphics[width=8cm]{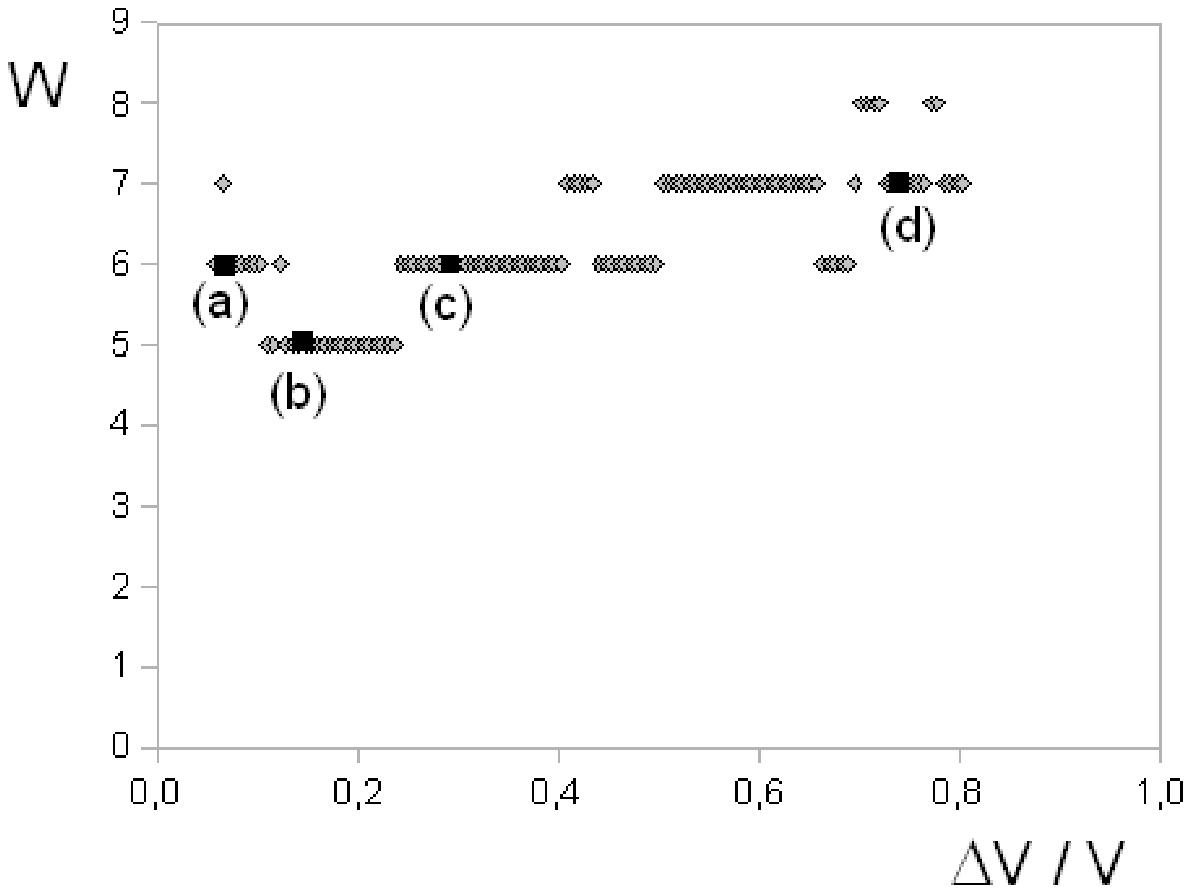}

\includegraphics[width=2cm]{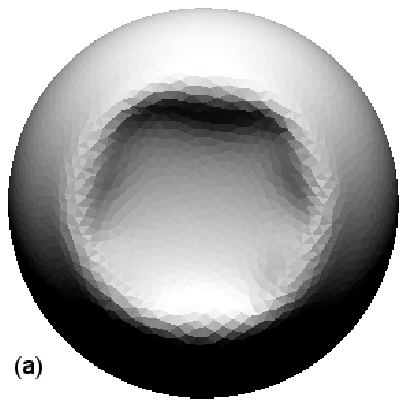}\includegraphics[width=2cm]{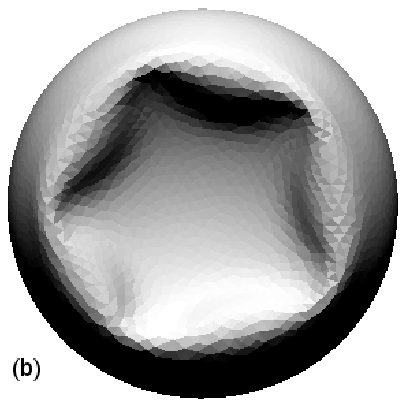}\includegraphics[width=2cm]{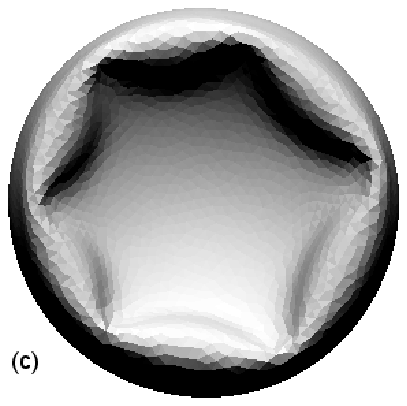}\includegraphics[width=2cm]{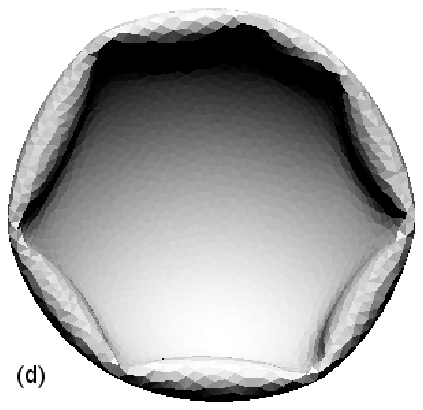}\end{center}

\caption{\label{fig:WrinklesNumber}Number of wrinkles (\emph{s}-cones) held
by the single depression after the secondary buckling of figure \ref{fig:DefEnergies}.
Poisson's ratio $\nu=0.5$; Föppl-von Karman number $\gamma=6.06\times10^{4}$($\frac{d}{R}=0.0122$).
Lower part: conformations at points indicated on the main figure.}

\end{figure}

\begin{figure}
\begin{center}\includegraphics[width=10cm]{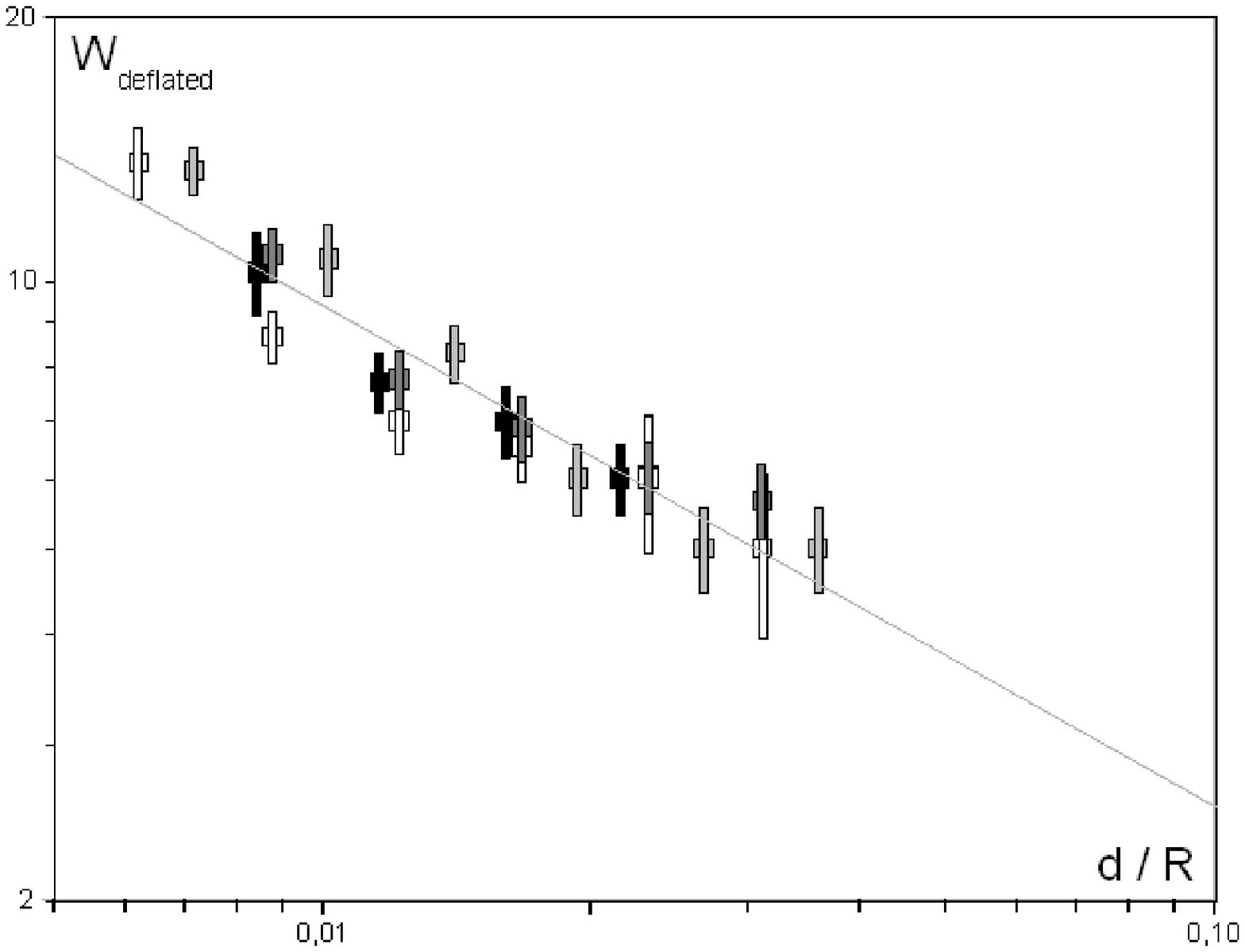}\end{center}

\caption{\label{fig:W_moyenn}Number $W_{deflated}$ of wrinkles (\emph{s}-cones)
at the end of a numerical deflation (averaged between $\frac{\Delta V}{V}=0.53$
and $\frac{\Delta V}{V}=0.76$). error bars are taken as the standard
deviation on this range, with a minimum value of $\pm0.5$. White:
$\nu=0.5$ ; light grey: $\nu=0$ ; dark grey: $\nu=-0.5$ ; black:
$\nu=-0.8$. Continuous line: $W_{deflated}=0.940\times\left(\frac{d}{R}\right)^{-\frac{1}{2}}$.}

\end{figure}

For the thinnest shells that undergo polygonal indentation, the most
conspicuous feature is the number $W$ of wrinkles, or \emph{s}-cones.
Figure \ref{fig:WrinklesNumber} shows the evolution of $W$ for a
typical deflation: first does $W$ decrease while the freshly nucleated
and still very flat depression hollows and enlarges, then it increases
again. Data are quite scattered: there is a typical noise of order
$\pm1$ on $W$, that has no observable correspondence in smooth energy
curves (fig. \ref{fig:buck2}). In order to decrease data scattering,
we calculated $W_{deflated}$ as the average value of $W$ between
$\frac{\Delta V}{V}=0.53$ and $\frac{\Delta V}{V}=0.76$ (these values
have been choosen in order to cover, for all the simulations, a significant
range of relative volume variation before autocontact, this latter
happening around $\frac{\Delta V}{V}\approx0.9$). Values of $W_{deflated}$
are comparable with results from the previous model, which did not
take gaussian curvature into account, and indicates a scaling law
in $\gamma^{-1/4}$ \cite{Quilliet 2008,JASA2011}. The dependence
of $W_{deflated}$ with $\gamma$ and $\nu$ is shown on figure \ref{fig:W_moyenn},
expressed in 3D parameters. It shows a scaling in $\left(\frac{d}{R}\right)^{-1/2}$,
which provides clues on the typical transversal size $l$ of the \emph{s}-cones
(presented on figure \ref{fig:SchemCaps}, left). Since \emph{s}-cones
stand alongside one another on a length which is of the order of an
equator, we can estimate $l$ as $\frac{2\pi R}{W_{deflated}}$. Hence
best fit of $W_{deflated}$ with dependence in$\left(\frac{d}{R}\right)^{-1/2}$
(figure \ref{fig:W_moyenn}) can be expressed as:\begin{equation}
l\approx6.7\,\sqrt{dR}\label{eq:TailleWrinkles}\end{equation}
\textcolor{black}{This result is fully comparable to the wrinkles
wavelength $4.7\sqrt{dR}$ that can be calculated from recent results
by Vella et al \cite{Vella2011} on the indentation of strongly pressurized
shells. As shown in reference \cite{Landau}, $\sqrt{dR}$ arises
naturally from balancing the bending and in-plane deformation energy
of a small deformation on a spherical shell. Recent results by}\cite{Knoche2011}\textcolor{black}{{}
showed that $\sqrt{dR}$ scales also for the rim width even in large
axisymmetric depressions ; we confirm here that it governs also other
types of large deformations such as} \emph{s}-cones transversal size
in polygonal depressions.

\section{Other postbuckling features}

First buckling and its consequences were presented in section \ref{sec:First-order-transition}.
For application purpose, we may interest to the inside/outside pressure
difference after its drop at first buckling. Careful examination of
the numerical data revealed supplementary features, uncorrelated with
the occurrence of the secondary buckling previously exposed : two
types of behaviour clearly appear in the evolution of the reduced
pressure $\frac{\Delta P}{\Delta P_{Landau}}$ during deflation.

For $\frac{d}{R}\gtrsim0.014$, pressure difference $\Delta P$ presents
the type of evolution calculated by \cite{Knoche2011}, \emph{i.e.}
quasi-plateauing after buckling (variation of about 15\% during the
whole deflation, plus some occasional dispersion due to numerical
procedure), up to autocontact. Furthermore, an order relation is respected:
at every volume step, the ratio $\frac{\Delta P}{\Delta P_{Landau}}$
weakly increases when $\gamma$ decreases. This is to be observed
in fig. \ref{fig:Pressure-difference-}, for the 4 curves corresponding
to the highest relative thicknesses.

For the thinnest shells ($\frac{d}{R}\lesssim0.012$ in the simulations
performed), $\frac{\Delta P}{\Delta P_{Landau}}$ regularly re-increases
with deflation after the pressure drop, crossing successively the
curves at smaller $\gamma$'s (as shown on the 2 {}``thinnest''
curves of fig. \ref{fig:Pressure-difference-}).

\begin{figure}
\begin{center}\includegraphics[width=10cm]{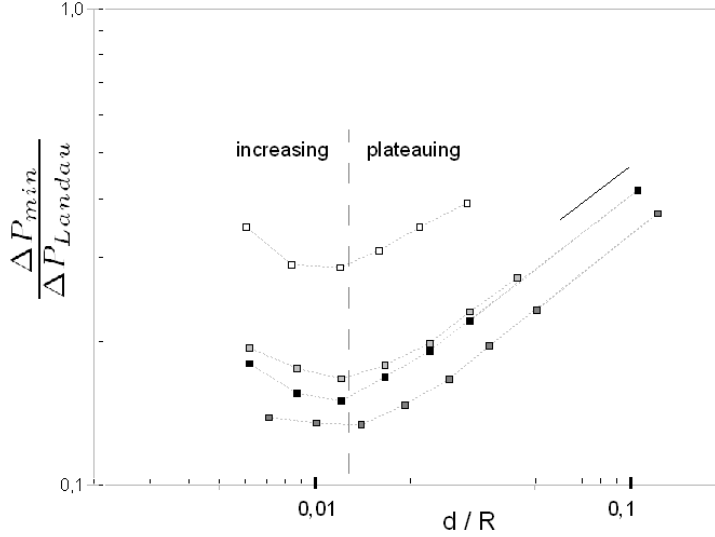}\end{center}

\caption{\label{fig:MinPressureForSeveralNus}Minimum inside/outside pressure
difference after buckling (cf figure \ref{fig:Pressure-difference-}),
adimensionalized by Landau pressure, logarithmic representation. Black:
$\nu=0.5$; dark grey: $\nu=0$; light grey: $\nu=-0.5$; white: $\nu=-0.8$.
Interrupted line separates the two types of evolution of the pressure
after the first buckling: increasing or plateauing (see figure\ref{fig:Pressure-difference-}
). Black line indicate slope 0.5.}

\end{figure}

\begin{figure}
\begin{center}\includegraphics[width=10cm]{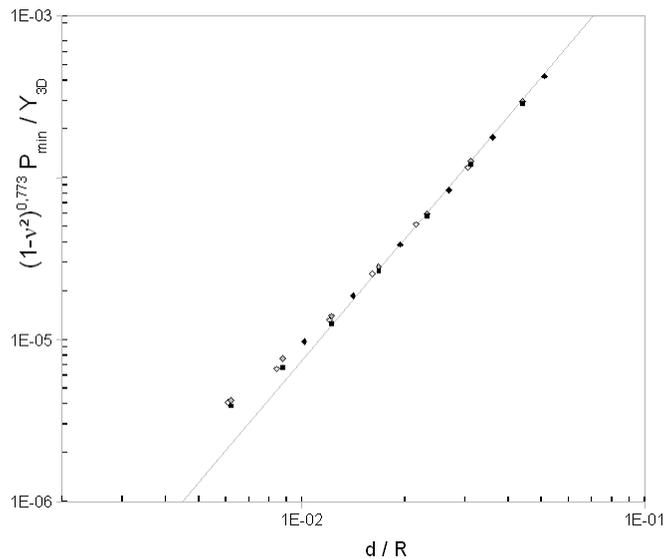}\end{center}

\caption{\label{fig:Master-curve}Pressure master curve after first buckling:
$\frac{\Delta P_{min}}{Y_{3D}}\times\left(1-\nu^{2}\right)^{0.773}$
versus $\frac{d}{R}$. Black squares: $\nu=0.5$; black diamonds:
$\nu=0$; light grey diamonds: $\nu=-0.5$; white diamonds: $\nu=-0.8$.
Continuous line: $\frac{\Delta P_{min}}{Y_{3D}}\times\left(1-\nu^{2}\right)^{0.773}=0.75\times\left(\frac{d}{R}\right)^{2.5}$.}

\end{figure}

In order to extract a general behaviour from these differents observations,
we focused on $\frac{\Delta P_{min}}{\Delta P_{Landau}}$, the minimum
value of $\frac{\Delta P}{\Delta P_{Landau}}$ after buckling. Figure
\ref{fig:MinPressureForSeveralNus} shows that the cross-over between
the two regimes around $\frac{d}{R}\approx0.014$ also corresponds,
for each $\nu$, to the minimum of relative pressure drop after the
first buckling. In the plateauing regime,$\frac{\Delta P_{min}}{\Delta P_{Landau}}$
stands for the plateauing value for $\frac{d}{R}\gtrsim0.014$ ; it
shows a power-law of the type $\frac{\Delta P_{min}}{\Delta P_{Landau}}=a\left(\nu\right)\times\left(\frac{d}{R}\right)^{0.5}$.
Similarly to what was done in section \ref{sec:First-order-transition},
and since $a\left(\nu\right)$ appears to be even (curves at $\nu=0.5$
and $\nu=-0.5$ almost mix up on figure \ref{fig:MinPressureForSeveralNus}),
we looked for a prefactor of the form $\left(1-\nu^{2}\right)^{x}$,
minimizing $x$ for the best fits at $\frac{d}{R}>0.014$. This led
us to propose the master curve presented on figure \ref{fig:Master-curve},
of formula $\frac{\Delta P_{min}}{Y_{3D}}\times\left(1-\nu^{2}\right)^{0.773}=0.75\times\left(\frac{d}{R}\right)^{2.5}$.
We do not have for the moment theoretical clues to justify these two
successive fitting operations, but (i) it allows to describe numerical
results in a very condensed way for $\frac{d}{R}\geq0.014$ , and
(ii) for all the shells these reduced values impressively gather on
a single curve, for $\gamma$ ranging from $8\times10^{2}$ to $4.7\times10^{5}$,
and for $\nu$ between -0.8 and 0.5. This result, exposed on figure
\ref{fig:Master-curve} using 3D parameters, is expected to be of
practical use for all experiments involving deflation controlled by
the volume. On a more conceptual point of view, plot clearly confirms
two different scalings of the pressure during deflation, around a
threshold in relative thickness $\left(\frac{d}{R}\right)_{c}\approx0.013$.
This may be an indication of the existence of different ways to accomodate
\emph{s}-cones on a sphere, and requires further investigations.

\section{Conclusion}

Systematic numerical study of the buckling of a spherical shell, in
the conformation with a single depression, allows to sketch the influence
of the different geometrical or elastic parameters through quite simple
theoretical or phenomenological laws. The surface model can be translated
in 3D parameters, that are the shell's thickness, and the two elastic
parameters of the material that compose it: Young modulus and Poisson's
ratio.

At imposed volume, the Young modulus does not play on the shape. Results
showed that the first transition (toward axysimmetrically buckled
shape), and the second one, with appearance of wrinkles, or {}``\emph{s}-cones'',
is mainly driven by $\frac{d}{R}$ for non-auxetic (\emph{i.e.} with
positive Poisson's ratio) materials. For auxetic materials, the Poisson's
ratio may have a determining importance, by strongly displacing transitions
toward higher values of the relative volume variation, up to possible
vanishing. Decreasing the Poisson's ratio down to very negative values
stabilizes spherical deflation at the expense of dimples creation,
and axisymmetric dimples against appearance of wrinkles.

The number of wrinkles indicate a dependence in $\left(\frac{d}{R}\right)^{-1/2}$,
that confirms $\sqrt{dR}$ as the accurate scaling for elastic deformations
of elastic spherical surfaces.

The Young modulus scales pressure features: critical inside/outside
pressure difference that triggers first buckling, and plateauing pressure
after buckling. Detailed behaviour, that is shown to reduce to a master
curve, opens the possibility for two different wrinkling regimes.

\section*{Acknowledgments}

The author thanks K. Brakke for developing and maintaining the Surface
Evolver software, including invaluable interactions during this work,
and P. Marmottant and F. Quéméneur for fruitful discussions.


\begin{thebibliography}{37}
\bibitem{NoteCourbure}\textcolor{black}{There may be discrepancies
between different communities about the definitions of both $\kappa$
and $c$. Here we used definitions that rule in the Soft Matter community,
and more particularly in physics of lipids vesicles: $c=\frac{1}{R_{1}}+\frac{1}{R_{2}}$,
where $R_{1}$ and $R_{2}$ are the main algebraic curvature radii
(mathematicians usually consider half of this quantity). And we take
$\kappa$ such that the bending energy of a sphere is $8\pi\kappa$.}
\bibitem{NoteKhi}A possible expression of the 2D compressibility
modulus is $\chi_{2D}=\frac{Y_{2D}}{2\left(1-\nu\right)}=\frac{Y_{3D}\times d}{2\left(1-\nu\right)}$
\bibitem{NoteChoixParamB2}Plotting $\ln\left\{ \left(\frac{\Delta V}{V}\right)_{buck\,2}\right\} $
versus $\ln\left(\frac{d}{R}\right)$, as for the first buckling,
excessively particularizes points at $\nu=-0.8$, which results in
a correlation coefficient of only -0.90.
\end{thebibliography}
\end{document}